\documentclass[a4paper,11pt]{scrartcl}
\usepackage{url,lineno,microtype,subcaption}
 \usepackage[urlcolor=blue,linkcolor=blue,citecolor=blue,colorlinks=true]{hyperref}
\usepackage[onehalfspacing]{setspace}
\usepackage{units}
\usepackage{journals}
\usepackage{color}
\usepackage{natbib}
\usepackage{amsmath}
\usepackage{graphicx}
\definecolor{WDcol}{rgb}{0.0,0.0,0.0}
\definecolor{WDcolb}{rgb}{0.0,0.0,0.0}

\newenvironment{keywords}%
{\begin{trivlist}\item[]{\bfseries\sffamily keywords:}\ }
{\end{trivlist}}

\newcommand\rev[1]{{\color{WDcol}#1}}
\newcommand\revb[1]{{\color{WDcolb}#1}}
\newcommand{\bs}[1]{\tilde{#1}}
\def\bm#1{\ensuremath{\mathchoice{\mbox{\boldmath$\displaystyle#1$}}
{\mbox{\boldmath$\textstyle#1$}}
{\mbox{\boldmath$\scriptstyle#1$}}
{\mbox{\boldmath$\scriptscriptstyle#1$}}}}

\author{W. Dietrich \& J. Wicht\\ {\scriptsize Max Planck Institute for Solar System Research, G\"ottingen, Germany}}
\title{Penetrative Convection in Partly Stratified Rapidly Rotating Spherical Shells}
\begin{document}

\maketitle

\begin{abstract}
Many celestial objects are thought to host interfaces between convective and stable stratified interior regions. \rev{The interaction between both, e.g. the transfer of heat, mass or angular momentum depends on whether and how flows penetrate into the stable layer.} Powered from the unstable, convective regions, radial flows can pierce into the stable region depending on their inertia (overshooting).  In rapidly rotating systems, the dynamics are strongly influenced by the Coriolis force and radial flows penetrate in stratified regions due to the geostrophic invariance of columnar convection even in the limit of vanishing inertia. Within this study\revb{,} we numerically investigate \rev{both mechanisms and hence explore the nature} of penetrative convection in rapidly rotating spherical shells. \rev{The study covers} a broad range of system parameters, such as the strength of the stratification relative to the Coriolis force or the inertia. Guided by the application to Saturn, we model a sandwiched stable stratified layer (SSL) surrounded by two convective zones. A comprehensive analysis of the damping behavior of convective flows at the edges of the SSL showed that the mean penetration depth is controlled by the ratio of stratified and unstratified buoyancy gradients and \rev{is hence} independent of rotation. A scaling law is derived and suggests \rev{that the penetration depth decreases with the square root of the ratio of \revb{unstabilizing} and \revb{stabilizing} entropy gradients}. The influence of the Coriolis force, however, is evident by a modulation of the penetration depth along latitude, since convective columns are elongated vertically and hence pierce predominantly into the SSL around mid-latitudes and outside the tangent cylinder. Our result also show that the penetration depth \rev{decreases linearly} with the flow length scale (\revb{low pass} filter), confirming predictions from the linear theory \rev{of rotating partially stratified convection}. 
\end{abstract}

\begin{keywords}
 stable stratification, rapidly rotating spherical shells, penetrative convection, numerical simulation, scaling laws
\end{keywords}

\section{Introduction}
When the local temperature gradient is steeper than the one associated with an adiabat, small perturbations from the hydrostatic equilibrium amplify to the \revb{well-known} Rayleigh-Taylor convective instability. This leads to vigorous convection \revb{that} very rapidly re-establishes bulk adiabatic gradients \rev{of} density and temperature due to the inherent mixing and heat transport efficiency.
 
However, stable stratified regions, \rev{in which} the heat flux is conductive or radiative, are \revb{widespread} phenomena in stars and planets. Those regions are \rev{caused} by either subadiabatic temperature or positive heavy element gradient. In stars, the large efficiency of radiative heat transport induces stratified regions, e.g. in the radiative part of the solar interior \citep[e.g.][]{Zahn1991,Brun2011}. Stable stratification due to temperature inversion or subadiabatic temperature lapse rates are known for the Gas Giants (Jupiter Galileo probe, \citep{Atkinson1996}) and outermost atmosphere of terrestrial planets, where radiation is most efficient. In the Earth's liquid outer core, the outermost layer seems stratified \revb{caused by} either a subadiabatic temperature gradient or a local enrichment of light elements \citep{Fearn1981, Lister1998}. Such a layer would have profound consequences for core convection and dynamo action in the core and be traceable in the geomagnetic field \citep{Nakagawa2011, Buffett2014}. A potential effect of electrically conducting, stable stratified layers surrounding a dynamo region are \revb{weak amplitude}, strongly axisymmetric surface magnetic fields and are often proposed  for Mercury's or Saturn's dynamo regions \citep{Stevenson1982,Schubert1988,Christensen2006b}. 
    
For Saturn a particular interesting scenario, consisting of a stratified layer between two convective ones is favored by the H/He demixing behavior \citep{Stevenson1977, Pustow2016,Schoettler2018} and in agreement with several observational constraints, such as the peculiarly axisymmetric magnetic field \citep[e.g.][]{Cao2012}, the apparently impeded thermal evolution \citep{Leconte2013,Nettelmann2013} and the detection of gravity wave induced pulsations in Saturn's rings \citep{Hedman2013,Fuller2014}. Saturn's interior might have been homogeneously mixed in the early, hot stages of the thermal evolution. Once the adiabatic temperature gradient crosses the demixing curve, He droplets form, sink downwards and are remixed at greater depth. This process \revb{builds} up a compositional gradient roughly around mid-depth bridging from He-depleted outer to He-enriched inner convective zone \citep{Stevenson1977,Schoettler2018}.

Radial, convective flows originating from the unstable regions might penetrate into the stratified region due \rev{to} inertia or the geostrophy of the columnar convective structures. It is useful \revb{to distinguish qualitatively} the different end-members scenarios \revb{in terms of} the dominant forces. In rotating, partially stratified convection, the three main forces, each associated with a \revb{timescale}, are the Coriolis force ($\tau_{rot}$), the buoyancy of the rising fluid parcel ($\tau_{buo}$) and the inverse buoyancy due to the stratification ($\tau_{strat}$). The time scales are given by the rotation rate $\Omega$, $\tau_{rot} = (2 \Omega)^{-1}$, the Brunt-V\"ais\"al\"a frequency $N$, $\tau_{strat} = \sqrt{N^{-2}} = (- g/ \rho \, d \rho /dr)^{-1/2} $ and the time scale of buoyancy to accelerate fluid parcels $\tau_{buo} = ( g/\rho \,  \rho^\prime / D)^{-1/2}$ , respectively. Here $\rho^\prime$ is the density fluctuation of \revb{a} buoyant fluid parcel\rev{, $\rho$ the ambient density, $D$ a typical length scale and $g$ the local gravity. For our setup, $\tau_{buo}$ characterizes the convective  and $\tau_{strat}$ the stratified regions.} The ratio of the time scales can be expressed as the convective Rossby number $Ro_c$ and the stratification parameter $I_s$:
\begin{equation}
Ro_c = \frac{\tau_{rot} }{\tau_{buo}} , \qquad I_s  = \frac{2\Omega}{N} =  \frac{\tau_{strat}}{\tau_{rot}}  \ .
\end{equation}

In slowly rotating systems, rotational forces are less important ($I_s < 1$, $Ro_c > 1$) and a hot fluid parcel is accelerated by the \rev{buoyancy} in the convective region, rises radially upwards and eventually pierces into the SSL, where the depth of penetration depends on the previously gained inertia. This represents the classical case of non-rotating penetrative convection, typically called \textit{overshooting} or \textit{inertia} penetration. For this end-member\revb{,} the penetration depth should be independent of colatitude, and may be the dominant form of penetration close to the rotation axis for $Ro_c<1$ and in the equatorial regions in our models.

In rapidly rotating systems\revb{,} convective flows are strongly modified by the rotation ($Ro_c < 1$). The dominant Coriolis force \rev{constrains} the flow to be invariant along the rotation axis leading to the \revb{well-known} convective columns and strongly geostrophic zonal flows. This implies that the flows extend into the stratified layer independent of inertia\rev{, but with a characteristic colatitudinal modulation}. When the Coriolis force is stronger than the gravity force associated with the stratification (i.e.~$I_s > 1$, $Ro_c < 1$),  \textit{rotational} penetration is strong and acts predominantly at mid-latitudes and outside the tangent cylinder (TC), where convective columns can vertically \rev{extend} through the whole spherical shell. Since \rev{in} the equatorial regions and close to the rotation axis no convective columns can be extended into the SSL, rotational penetration yields a characteristic colatitude dependence. If, on the other hand, $I_s < 1$ and $Ro_c < 1$ the SSL should be devoid of radial flows since the stratification is strong enough to efficiently break the vertical stiffness of geostrophic flows.

\rev{A fundamental understanding of penetrative convection in rapidly rotating spherical shells is of ample importance in geo- and astrophysics.} The efficiency or vigor of penetrative convection controls the transport of heat\rev{,} mass, angular momentum and magnetic fields across the interface between unstable \rev{and} stable stratif\rev{ied regions}. If the penetration is strong and \revb{the associated} heat transport efficient, adiabatic regions are extended into the stratified layers \citep[e.g.][]{Browning2004}. This means that stratified layers can be eroded by the permanent entrainment of convective flows and subsequent efficient mixing \rev{ \citep[e.g.][]{Ellison1959,Levy2002}. Another complication arises from \revb{the combination of} spherical geometry and rapid rotation.} If rotational forces dominate over the inverse buoyancy associated with the stratification, piercing radial flows are \rev{non-uniform along colatitude} leading to latitudinal entropy (density) gradients and hence drive baroclinic instabilities. \rev{Those in turn will act as a boundary condition for the differential rotation in the convective regions and potentially alter the magnetic field \citep{Stevenson1982}. }

Classic studies of penetrative convection dealt with non-rotating cartesian setups of Rayleigh-B\'enard convection \citep{Veronis1963}. The effect of rotation was firstly considered in the framework of oceanic dynamics \citep[e.g.][]{Julien1996}. The investigation of rotating convection in spherical shells below or above a stable stratified layer was mainly driven by studies on the interaction and efficiency of radiative and/or convective heat transport in solar or stellar interiors \citep[e.g.][]{Zahn1991, Brummell2002,Rogers2006}. For rapidly rotating A-type stars, where a convective core is surrounded by a radiative outer envelope, \citet{Browning2004} reported that rotational and overshooting penetrative convection generates adiabatic regions in the radiative zone preferentially at higher latitudes (prolate adiabatic core).

Several numerical investigations tuned to the solar setup with a convective envelope enclosing a deeper radiation dominated zone, investigate the parameter dependence of penetrative convection  \citep{Zahn1991, Hurlburt1994,Rogers2006}. Since the stratification originates from enhanced radiative heat transport, it is typically assumed that penetration depth is largely characterized by two parameters, one being the P\'eclet number given by $Pe=U\,D/\kappa$, where $U$ is a typical flow speed, $D$ a length scale and $\kappa$ the thermal diffusivity measuring the radiative heat transport \citep{Zahn1991,Brummell2002}. The other parameter, $S$ is the ratio of sub- and superadiabaticity in the stable and unstable region \citep{Hurlburt1994}.  However, different scaling relations between the penetration depth and $S$ have been reported \citep{Hurlburt1994,Brummell2002,Rogers2006}. \rev{Interestingly, \revb{in models that take} spherical geometry and rotation into account, the associated Coriolis forces appear \revb{unimportant} for the \revb{magnitude of the penetration} depth.}

In the planetary context, the Galileo Probe revealed subadiabatic temperature gradients and non-ceasing zonal flows in Jupiter's outermost atmosphere down to roughly 20 bar of atmospheric pressure. \rev{This suggests that zonal flows are not damped, but \revb{rather} maintained in stratified regions \revb{yet} disconnected from their potential sources, such as convection or irradiation gradients. Assuming that the axisymmetric winds are driven from correlations of the deep convection (i.e. Reynolds stresses), \citet{Zhang1996,Zhang1997} showed that zonal mean flows are} vertically extended by the Coriolis forces in accordance with the Taylor-Proudman theorem\rev{, but non-axisymmetric convective flows are damped by the inverse buoyancy gradients due to \revb{the}  stratification.}  From a more theoretical point of view, the \rev{linearized} model of inviscid, inertia-less, rapidly rotating penetrative convection as studied by \citet{Takehiro2001} show\rev{s} a \textit{linear} length scale dependence of penetration depth such that larger scale flows (like differential rotation) are less damped than short length scale flows typically associated with convective motions. The overall damping amplitude depends on the gravity force associated with the stable stratification relative to the Coriolis force ($I_s$). Subsequent studies including nonlinear inertia showed a more complex damping behavior \citep{Takehiro2002}, but the results offered \revb{an} attractive explanation how zonal flows are extended through stratified regions. 

\rev{The major aim of the present study is to understand how overshoot and rotational penetration act across the interface between stratified and unstratified regions in rapidly rotating, spherical shell models, which are most suitable for interiors of planets. Even though the effects of stratification on differential rotation and magnetic fields as far as the emerging waves inside the SSL are left aside \revb{in this study}, the investigation of fundamental properties of rapidly rotating, penetrative convection serves as a basis of upcoming research.} Our models cover a comprehensive part of the parameter space, such that the different regimes appropriate for planets \revb{are} reached in terms of $I_s$ and $Ro_c$. The general setup featuring a thin\revb{,} sandwiched stratified layer centered at \rev{mid-depth} and surrounded by \revb{two} thick convective zones is motivated from the application to Saturn.

\section{Model}
\begin{figure}
\centering
\includegraphics[width=0.9\columnwidth]{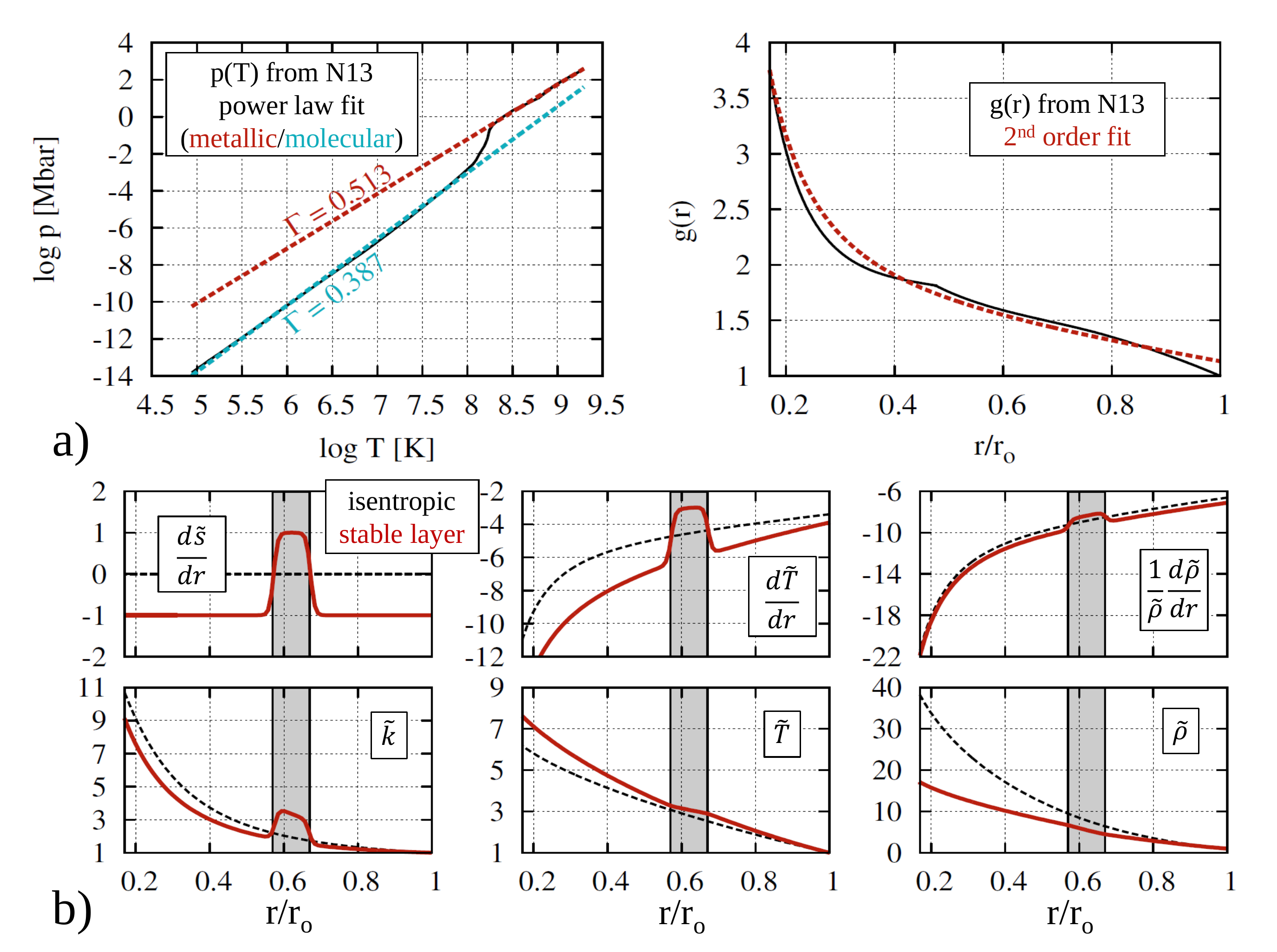}
\caption{\rev{a) power law ($\Gamma$) and 2nd order polynomial (gravity) fits to the interior state model from \citet{Nettelmann2013} b) the non-adiabatic background state is characterized by a prescribed entropy gradient profile ($d\bs{s}/dr$) setting the temperature, density and thermal conductivity. For visualization purposes the relative non-adiabaticity is strongly exaggerated (parameters $\epsilon_s = 0.5$, $A_{SSL} = 1.0$, $Di=3.0$, $\Gamma=0.513$). }}
\label{figbackground}
\end{figure}

The non-dimensional governing equations for conservation of mass, momentum and thermal energy for an ideal gas in the anelastic approximation are given by \rev{\citep{Jones2011, Gastine2012, Verhoeven2015, Wicht2018}}:

\begin{align}
\bm{\nabla} \cdot \bs{\rho} \bm{u} &= 0 \\
\frac{\partial \bm{u}}{\partial t} + \bm{u} \cdot \bm{\nabla} \bm{u} +\frac{2}{E} \bm{e}_z \times \bm{u} &= -\bm{\nabla} p^\star  + \frac{Ra}{Pr}  g s^\prime \bm{e}_r + \bm{F}_\nu \label{eqnstfin}\\
\frac{\partial s^\prime}{\partial t} + \bm{u} \cdot\bm{\nabla} s^\prime + u_r \frac{d \bs{s}}{dr}   &= \frac{1}{Pr \bs{\rho} \bs{T}} \bm{\nabla} \cdot \left( \bs{\kappa} \bs{\rho} \bs{T} \bm{\nabla}   s^\prime   \right) + \frac{Pr \, Di}{Ra} \frac{1}{\bs{\rho} \bs{T}}  Q_\nu \label{eqheatfin}\ ,
\end{align}
where $p^\star$ is the reduced pressure, $\textbf{F}_\nu$ is the viscous force, 
\begin{equation}
 \textbf{F}_\nu = \frac{1}{\bar{\rho}} \left[ \frac{\partial}{\partial x_j}
\bar{\rho} \left( \frac{\partial u_i}{\partial x_j} +
\frac{\partial u_j}{\partial x_i} \right) -
 \frac{2}{3 } \frac{\partial}{\partial x_i} \bar{\rho}
\frac{\partial u_j}{\partial x_j} \right] \\
\end{equation}
and $Q_\nu$ represents viscous heating given by 
\begin{equation}
Q_\nu =\sigma_{ij}\frac{\partial u_i}{\partial x_j} \, \text{,} \quad 
\sigma_{ij}=\bar{\rho}\left(\frac{\partial u_i}{\partial x_j}+\frac{\partial u_j}{\partial x_i}-\frac{2}{3}\delta_{ij}\nabla\cdot{\bm u}\right) \ ,
\end{equation}
where $\sigma_{ij}$ is the stress tensor. The non-dimensional parameters are given by
\begin{equation}
Ra  = \frac{\alpha_o T_o g_o d^4}{c_p \nu \kappa_0} \left\vert \frac{d\bs{s}}{d r} \right\vert_{r_i} \, \text{,} \quad
Pr = \frac{\nu}{\kappa_o} \, \text{,} \quad
E = \frac{\nu}{\Omega d^2} \ .
\end{equation}
These equations were formulated for an adiabatic background state. In our model, a prescribed non-adiabaticity of the background state powers or suppresses convection. The amplitude of these \revb{destabilizing} and \revb{stabilizing} entropy gradients must remain small relative to the adiabat to make sure the equations still hold. The non-adiabaticity of the new background state is thus scaled  with $\epsilon_s \ll 1 $, 

\begin{equation}
\frac{d\bs{T}}{dr} = \epsilon_s \bs{T} \frac{d\bs{s}}{dr} - Di \, \bs{g}(r) 
 \, \text{,} \qquad
\frac{1}{\bs{\rho}} \frac{d \bs{\rho}}{dr} =  \epsilon_s \frac{d\bs{s}}{dr} - \frac{Di}{\Gamma \bs{T} } \bs{g}(r) \ ,
\label{eqdefgrarhond}
\end{equation}
where $d\bs{s}/dr$ is the dimensionless analytically prescribed stratification profile and \rev{$\Gamma$ the Gr\"uneisenparameter which is in an ideal gas related to the polytropic index $n$ and specific heats by
\begin{equation}
\Gamma = \frac{c_p}{c_v} - 1 = \frac{1}{n} \ .
\end{equation}  } The background state is \revb{characterized} by \rev{the relative deviation from an adiabat $\epsilon_s$ and the dissipation number $Di$, which sets the background density variation}
\begin{equation}
\epsilon_s  = \frac{d}{c_p} \left\vert \frac{d\bs{s}}{dr} \right\vert_{r_i} \ \text{,} \quad Di = \frac{\alpha_o g_o d}{c_p}\ ,  
\end{equation}
where $\vert d\bs{s} /dr \vert_{r_i}$ is the dimensional reference entropy gradient at the inner boundary. The background entropy gradient is prescribed analytically and sets convective (unstable stratified, $d\bs{s}/dr < 0$) and stable ($d\bs{s}/dr > 0$) regions. The $d\bs{s}/dr$-profile is given by
 \begin{equation}
 \frac{d\bs{s}}{dr}  =\left( \frac{  A_{SSL}+1 }{4} \right) \left( 1+ \tanh\Big[ (r-r_{lb})\, d_s \Big] \right) \left(1- \tanh\Big[(r-r_{ub})\, d_s \Big] \right) -1 \label{eqdefentropy0} \ ,
\end{equation}
where $A_{SSL}$ is the amplitude of stable stratification (basically setting the BV frequency\rev{, see eq.~\ref{eqnondimno}}), $r_{lb}$ and $r_{ub}$ are lower and upper boundary of the SSL. The parameter $d_s$ defines the slope of the profile. When choosing the stratification amplitude, $\epsilon_s \, A_{SSL} \ll 1$ is required to be compliant with the treatment in the framework of the anelastic approximation, which is based on an adiabatic and steady background state. The gravity profile and $\Gamma$ are fitted to an interior model of Saturn by \citet{Nettelmann2013} (see also fig.~\ref{figbackground}), \rev{where a polynomial of second order is used for the former}:
\begin{equation}
\bs{g}(r) = g_0 + g_1 \frac{r}{r_o} + g_2 \frac{r_o^2}{r^2} \ , 
\label{eqgrav}
\end{equation}
with 
\begin{equation}
g_0 = 1.854 \quad \& \quad g_1 = -0.781 \quad \& \quad g_2 = 0.0558 \ .
\end{equation}
$\Gamma$ characterizes the nature of the ideal gas and is found by fitting $p$-$T$-curves from \citet{Nettelmann2013} with polytropic laws such that
\begin{equation}
p \propto T ^{\frac{1+\Gamma}{\Gamma}} \ ,
\end{equation}
yielding a best fitting $\Gamma=0.513$ for the deep interior. \rev{The mean density contrast is set by $Di=3$, corresponding to \revb{a} top-to-bottom ratio of $\rho_i/\rho_o\approx 38$. Then the temperature and density gradient are set by a chosen entropy gradient profile and eq.~\ref{eqdefgrarhond}. Fig.~\ref{figbackground}, panel b) shows the so derived background state, indicating sub- and superadiabatic temperature gradients (middle panel, upper row). The black, dashed line corresponds to an isentropic ($d\bs{s}/dr=0$) model for reference. Note, that the non-adiabaticity in the figure is strongly exaggerated using $\epsilon_s=0.5$ and $A_{SSL}=1.0$. For production runs, the relative deviation from an adiabatic background must remain small ($\epsilon_s A_{SSL} \ll 1$). } To ensure that the newly defined background state is steady, the heat flux must be constant on every radius. Hence a thermal conductivity profile ($\bs{k}=\bs{\rho} c_p \bs{\kappa}$) is set such that
\begin{equation}
\label{eqdefkappa}
\bs{k} = \frac{Q_0}{ 4 \pi} \left( r^2 \frac{d\bs{T}}{dr} \right)^{-1}  \ ,
\end{equation}
where $Q_o$ is the heat flux at the outer boundary. \rev{The profile of $\bs{k}$ is shown in fig.~\ref{figbackground}, panel b) bottom left plot.}

The equations are \revb{non-dimensionalized} by scaling length with the shell thickness $d = r_o-r_i$, time by the viscous diffusion time $\tau_\nu = d^2/\nu$ and entropy by $d \vert d\bs{s}/dr \vert_{r_i}$. To avoid radius dependent control parameters, reference values for $\bs{\rho}$, $\bs{T}$, $\bs{\kappa}$, $\bs{g}$ are taken at the outer boundary. The convective Rossby number $Ro_c$, \revb{the} non-dimensional Brunt-V\"ais\"al\"a frequency and the non-dimensional stratification parameter are 
\begin{equation}
Ro_c = \sqrt{\frac{Ra}{Pr}} E \, \text{,} \quad N =\text{max}\left[ \sqrt{\frac{Ra}{Pr} \bs{g}(r) \frac{d\bs{s}}{dr}} \right]\, \text{,} \quad
I_s =\text{max} \left[ \frac{2}{Ro_c} \left( \bs{g}(r) \frac{d\bs{s}}{dr}\right)^{-1/2} \right] \ \label{eqnondimno}.
\end{equation}

\section{Numerical results}
Eqs.~\ref{eqnstfin} and \ref{eqheatfin} are solved numerically using MagIC 5.6 \citep{Wicht2002,Gastine2012, Schaeffer2013} which is modified such that the non-adiabatic background state relations (eq.~\ref{eqdefgrarhond}) are integrated. The mechanical confinement yields impenetrable and free-slip conditions at both walls. The entropy boundary conditions are \revb{fixed-flux} in accordance with the stratification profile (eq.~\ref{eqdefentropy0})\revb{, which sets} unstable and stable regions. The stratified \revb{layer} is located between $r_{lb}/r_o= 0.57$ and  $r_{ub}/r_o=0.67$. The numerical grid resolution is $N_r \times N_\vartheta \times N_\phi = 145 \times 256 \times 512$ for $E=10^{-4}$, $193 \times 320 \times 640$ at $E=3 \cdot 10^{-5}$ and $241 \times 320 \times 640$ at $E=10^{-5}$. The spectral resolution is limited to $N_\ell = 2/3\, N_\vartheta$.  Fixed parameter values are the aspect ratio $\beta=r_i/r_o = 0.17$, $\epsilon_s = 10^{-4}$, $d_s = 75$, gravity according to eq.~\ref{eqgrav}, $Di=3$ and $\Gamma = 0.513$.

\begin{table*}[!ht]
\centering
\renewcommand{\arraystretch}{1.1} 
\begin{tabular}{c|cccccccc}
no. & Ra & Pr & E &  $A_{SSL}$  &$I_s$ & $Ro_c$ & $\delta_r^{+}$ & $\delta_r^{-}$\\
 \hline
1.1 & $3 \cdot 10^7$ & 0.5 & $10^{-4}$& 500 & 0.09 & 0.775  & $7.59 \cdot 10^{-3}$ & $7.68 \cdot10^{-3}$  \\
1.2  & $3 \cdot 10^7$  & 0.5 & $10^{-4}$ & 300 &0.118 & $0.775$  & $8.24 \cdot 10^{-3}$ & $8.35 \cdot 10^{-3}$\\
1.3 & $3 \cdot 10^7$ & 0.5 & $10^{-4}$& 100 & 0.205 & 0.775  & 0.0130 & 0.0111  \\
1.4 & $3 \cdot 10^7$ & 0.5 & $10^{-4}$& 30  & 0.373 & 0.775 & 0.0167 & 0.0162 \\
1.5 & $3 \cdot 10^7$ & 0.5 & $10^{-4}$& 10  & 0.647 & 0.775 & 0.0314 & 0.0247 \\
1.6 & $3 \cdot 10^7$ & 0.5 & $10^{-4}$& 3   & 1.18  & 0.775  & 0.0612 & 0.0533  \\
1.7 & $3 \cdot 10^7$ & 0.5 & $10^{-4}$& 1   & 2.05  & 0.775  & 0.144 & 0.107  \\
1.8 & $3 \cdot 10^7$ & 0.5 & $10^{-4}$& 0.3 & 3.73  & 0.775  & 0.377 &0.221  \\
1.9 & $3 \cdot 10^7$ & 0.5 & $10^{-4}$& 0.1 & 6.47  & 0.775  & 0.688 & 0.238  \\
1.10 & $3 \cdot 10^7$ & 0.5 & $10^{-4}$& 0   & $\infty$   & 0.775   & 0.805 & 0.2446\\
\hline
2.1 & $7 \cdot 10^5$ & 0.5 & $10^{-4}$& 300 & 1.03  & 0.1 & 0.0173 & 0.0721 \\
2.2 & $1 \cdot 10^6$ & 0.5 & $10^{-4}$& 300 & 0.63  & 0.141 & 0.0136 & 0.0561  \\
2.3 & $3 \cdot 10^6$ & 0.5 & $10^{-4}$& 300 & 0.373 &  \rev{0.245} & 0.0166 & 0.0266  \\
2.4 & $1 \cdot 10^7$ & 0.5 & $10^{-4}$& 300 & 0.205 & 0.447 &   $8.095 \cdot 10^{-3}$ & $7.01 \cdot 10^{-3}$ \\
2.5 & $5 \cdot 10^7$ & 0.5 & $10^{-4}$& 300 & 0.092 & 1.0 &  $8.875 \cdot 10^{-3}$  & $8.40 \cdot 10^{-3}$ \\
2.6 & $7 \cdot 10^7$ & 0.5 & $10^{-4}$& 300 & 0.077 & 1.18 &   $8.18 \cdot 10^{-3}$ & $8.055 \cdot 10^{-3}$ \\
2.7 & $1 \cdot 10^8$ & 0.5 & $10^{-4}$& 300 & 0.065 & 1.41 &   $7.83 \cdot 10^{-3}$  & $6.77 \cdot 10^{-3}$  \\
\hline
3.1 & $1 \cdot 10^7$   & 0.3  & $10^{-4}$& 500 & 0.119& 0.574 & $8.02 \cdot 10^{-3}$ & $7.585 \cdot 10^{-3}$ \\
3.2 & $5.5 \cdot 10^7$ & 1.0  & $10^{-4}$& 330 & 0.118& 0.742 &$6.68 \cdot 10^{-3}$ & $7.51 \cdot 10^{-3}$ \\
3.3 & $5.5 \cdot 10^7$ & 0.75 & $10^{-4}$& 250 & 0.118& 0.854 & $7.59 \cdot 10^{-3}$ & $6.79 \cdot 10^{-3}$\\
3.4 & $3 \cdot 10^7$   & 0.3  & $10^{-4}$& 200 & 0.117& 1 & $0.0107$ & $8.84 \cdot 10^{-3}$  \\
3.5 & $5.5 \cdot 10^7$ & 0.3  & $10^{-4}$& 100 & 0.118& 1.352 & 0.0106 & $8.48 \cdot 10^{-3}$\\
\hline
4.1 & $3 \cdot 10^7$ & 0.5  & $10^{-4}$ & 75  & 0.236 & 0.775   & 0.0132 & 0.0120 \\
4.2 & $3 \cdot 10^7$ & 0.5  & $10^{-4}$ & 7.5  & 0.746 & 0.775  & 0.035 & 0.0292 \\
4.4 & $3 \cdot 10^7$ & 0.5  & $10^{-4}$ & 0.75  & 2.36 & 0.775   & 0.1748 & 0.1339 \\
4.5 & $1.1 \cdot 10^8$ & 1.0 & $3 \cdot 10^{-5}$ & 500 & 0.236 & 0.316 & $6.44 \cdot 10^{-3}$ & $7.32 \cdot 10^{-3}$ \\
4.6 & $1.1 \cdot 10^8$ & 1.0 & $3 \cdot 10^{-5}$ & 50 &  0.746 & 0.316 & 0.0145 & 0.0145\\
4.7 & $1.1 \cdot 10^8$ & 1.0 & $3 \cdot 10^{-5}$ & 5 & 2.36 & 0.316 & 0.0421 & 0.0310\\
4.8 & $1.1 \cdot 10^8$ & 1.0 & $3 \cdot 10^{-5}$ & 1 & 5.277 & 0.316 & 0.143 & 0.1062\\
4.9 & $6.3 \cdot 10^8$ & 1.0  & $10^{-5}$ & 750 &  0.236 & 0.245 & $6.22\cdot 10^{-3}$ & $9.18 \cdot 10^{-3}$ \\
4.10 & $6.3 \cdot 10^8$ & 1.0  & $10^{-5}$ & 75 &  0.746 & 0.245 & 0.0117 & 0.0161\\
4.11 & $6.3 \cdot 10^8$ & 1.0  & $10^{-5}$ & 7.5 & 2.36 & 0.245 & 0.0319 & 0.0309\\
4.12 & $6.3 \cdot 10^8$ & 1.0  & $10^{-5}$ & 1.0 & 6.463 & 0.245 & 0.1593 & 0.1117\\
\hline
\end{tabular}
\caption{Numerical models performed. Fig.~\ref{figregime}, left plot scatters the models over their associated time scale ratios, i.e. $Ro_c$ and $I_s$.}
\label{tabruns}
\end{table*}

\begin{figure} 
\centering
\includegraphics[width=0.6\columnwidth]{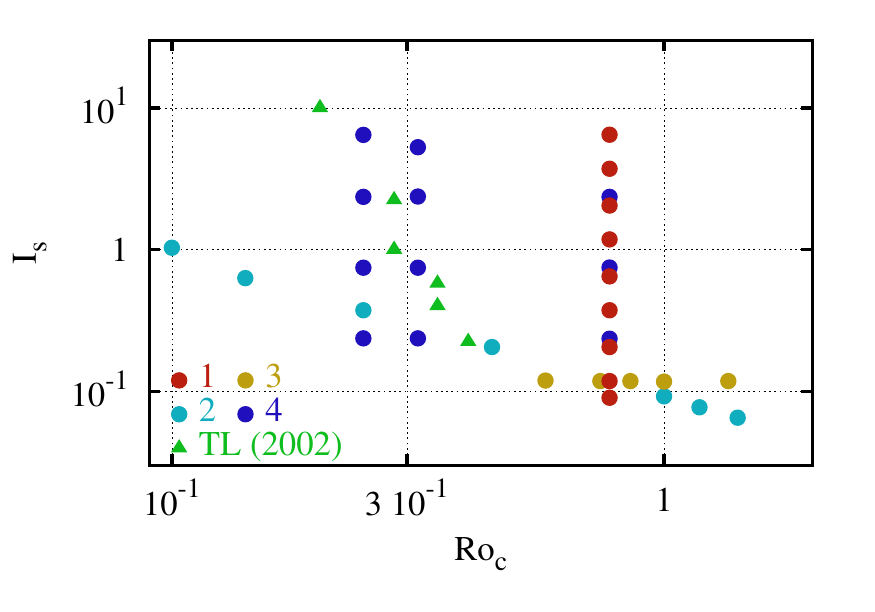}
\caption{Regime diagram \revb{of the numerical models} from tab.~\ref{tabruns} sorted by their respective convective Rossby \revb{($Ro_c$)} and the stratification relative to rotation rate \revb{($I_s$)}. The colors \revb{indicate} simulations from different groups. The green triangles mark the cases used in \citet{Takehiro2002}. }
\label{figregime}
\end{figure}

\begin{figure}
\centering
\includegraphics[width=0.9\columnwidth]{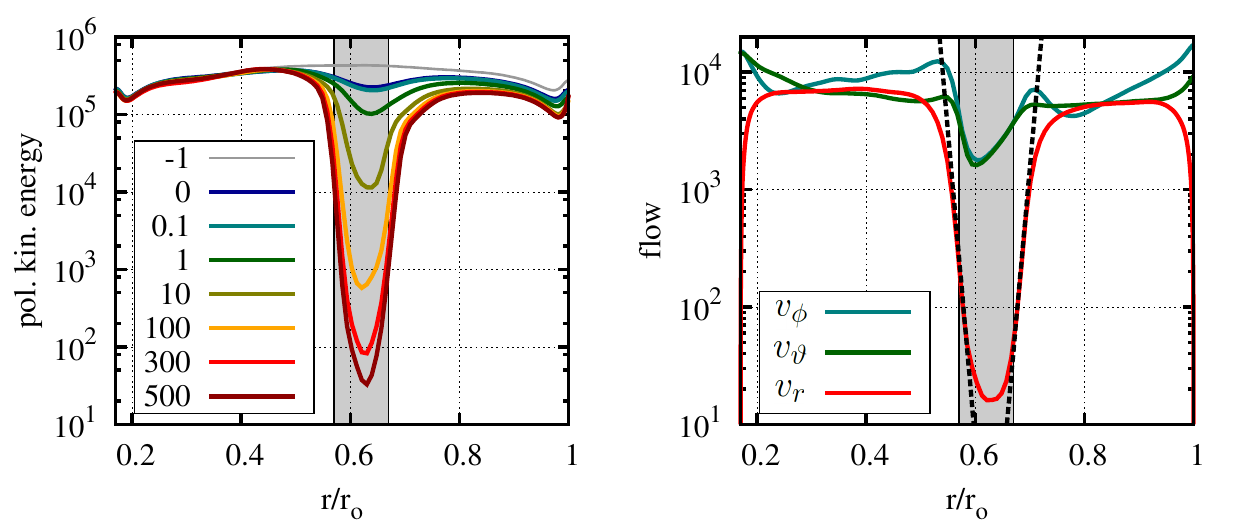}
\caption{Time averaged radial profiles of the non-axisymmetric poloidal energy for various stratification strength $A_{SSL}$ \revb{(left)} and horizontally averaged intensity of non-axisymmetric flow components $v_r$, $v_\phi$, $v_\vartheta$ \revb{for model 1.2 with $A_{SSL} = 300$} \revb{(right)}. \revb{The} dashed lines \revb{denote the assumed exponential decay laws}.  }
\label{figprofiles}
\end{figure}

\begin{figure}
\centering
\includegraphics[width=0.9\columnwidth]{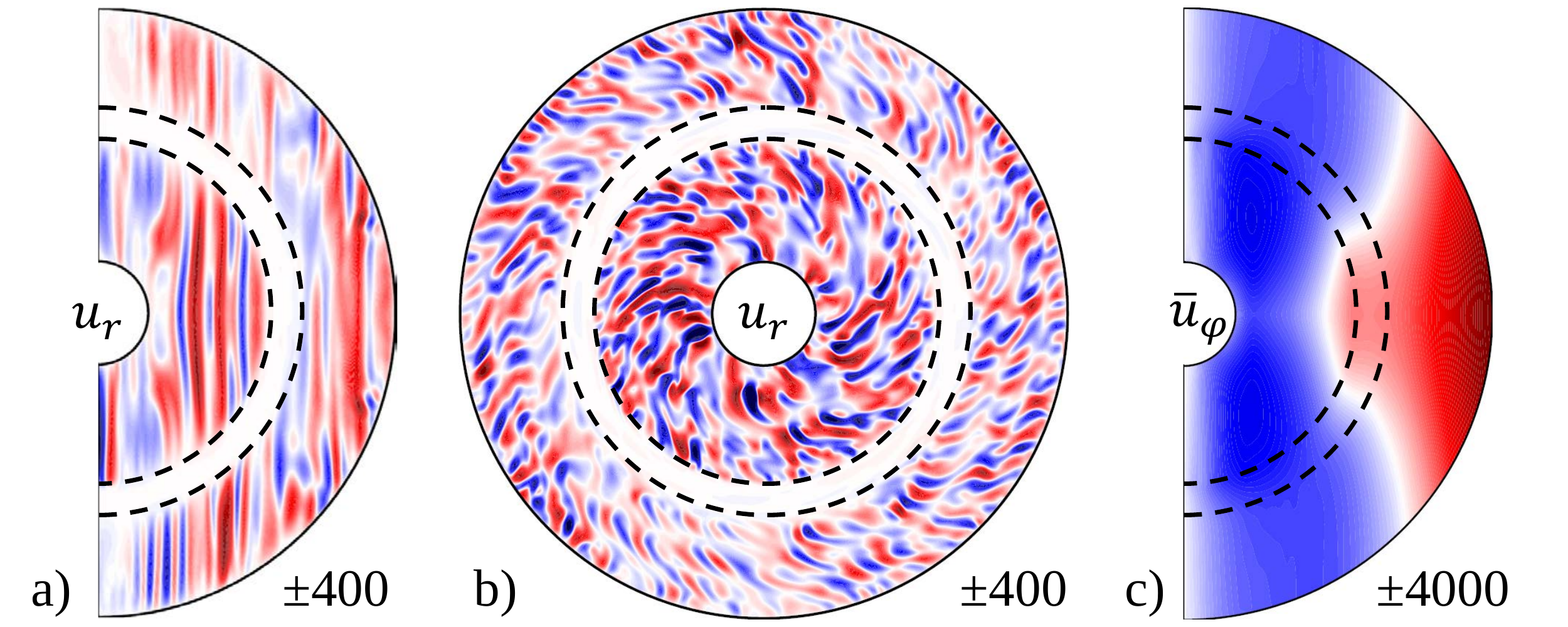}
\caption{Instantaneous radial flow in a meridional slice (a) and in the equatorial plane (b) as far as the axisymmetric zonal flow averaged over time (c) for model 4.10 from tab.~\ref{tabruns}\revb{.}}
\label{figc410_planes}
\end{figure}

The other parameters, $Ra$, $E$, $Pr$, $A_{SSL}$ are \revb{varied} to explore the parameter \revb{dependencies}. Fig.~\ref{figregime} provides an overview of $Ro_c$ and $I_s$ for the four sets of models. In models from group 1 the stratification amplitude (hence $I_s$) is varied and $Ro_c$ kept fixed. Group 2 investigates the influence of more vigorous convection, where increasing $Ra$ yields \revb{a decrease of} $I_s$, but \revb{an increase} of $Ro_c$.  For group 3, $I_s$ is kept constant, but with different combinations of $Pr$, $Ra$ and $A_{SSL}$. Finally\revb{,} group 4 investigates the Ekman number ($E$) dependence.

The immediate effect of adding a stratified layer can be seen in the radial profiles of non-axisymmetric poloidal kinetic energy for models from group 1 displayed in fig.~\ref{figprofiles}. For instance, the dark-red profile indicates that the non-axisymmetric poloidal kinetic energy \revb{is} suppressed by up to four orders of magnitude relative to a model without stratification (grey). Along both edges of the SSL region (highlighted in grey)\revb{,} the kinetic energy drops sharply. The inverse buoyancy in the SSL also reduces the flow amplitude outside of the stable zone. This might be linked to the columnar structures of the convective flows since their vertical extension along the rotation axis is limited by an increasing stratification in the SSL. \revb{In addition} an asymmetry between the lower and upper interface is clearly visible when the stratification is stronger, where for the upper one the radial flows seem to require more buoyancy to replenish after being reflected at the SSL underneath. This might be due to the fact, that the SSL itself acts as solid bottom wall hence providing a virtual tangent cylinder at $r_{ub}/r_o=0.67$.

\rev{The inverse buoyancy force in stratified regions directly acts to damp radial flows, which are consequently diverged into horizontal directions. Hence\revb{,} the effect of the SSL should be clearly visible in each of the non-axisymmetric flow components. E.g., is the time-averaged intensity of radial flows $v_r$ given by:
\begin{equation}
v_r (r,\vartheta) = \frac{1}{2\pi r^2} \sqrt{\langle {u_r^\prime}^2 \rangle_{t,\phi}} \ ,
\end{equation}
where the prime indicates non-axisymmetric flow. $v_\vartheta$ and $v_\phi$ are defined accordingly. Fig.~\ref{figprofiles}, b) shows the radial profiles of $v_r$, $v_\phi$ and $ v_\vartheta$ for model 1.2 (red profile in panel a). It \revb{is obvious}, that the poloidal energy is a rather good \revb{proxy} of the radial flow intensity. Since the radial flows are mainly deflected into horizontal directions, $v_\phi$ and $v_\vartheta$ appear muss less damped inside the SSL. Those represent wave-like, horizontal flows with frequencies smaller than the rotation or Brunt-V\"ais\"al\"a frequency. The strong decay of $v_r$ allows to investigate the damping behavior. \revb{For guidance,} the dashed lines in the figure represent exponential functions suggestive of exponential damping at the edges of the stratified regions\revb{. Their decay exponents are the penetration depths from tab.~\ref{tabruns}. The actual determination of the penetration depth is more involved and discussed in sec.~\ref{secpendepth} and fig.~\ref{figtest_pen}.} }

Those models might be biased by the choice of a somewhat large Ekman number (too viscous). For a more detailed inspection, a model with smaller $Ro_c$ hence stronger rotational constraints is favored (model 4.10). Fig.~\ref{figc410_planes} shows the instantaneous radial flow along a meridional cut (a) and in the equatorial plane (b). For this particular model, $I_s=0.746$ and $Ro_c=0.245$. It is obvious that the stratification breaks the geostrophy of the convective columns and efficiently wipes out radial flows in the SSL. This shows that in the SSL the inverse buoyancy exceeds the Coriolis force ($I_s < 1$). For both convective regions, the vertical length scale is clearly larger than the horizontal ones showing the effect of $Ro_c<1$. Further, the convective structures outside the tangent cylinders given by $r_i/r_o$ and $r_{ub}/r_o$, respectively, are vertically more extended than the corresponding convective flows inside the TCs. \rev{Furthermore\revb{,} fig.~\ref{figc410_planes}, panel c) indicates that the zonal flow is connected through the SSL between the convective shells. At the outer boundary of the model domain representing the planetary surface, the equatorial region features a wide prograde jet reminiscent of Saturn's equatorial \revb{super-rotating} jet}.

\subsection{radial flows in the vicinity of stratified layers}

\begin{figure*}
\centering
\includegraphics[width=0.9\textwidth]{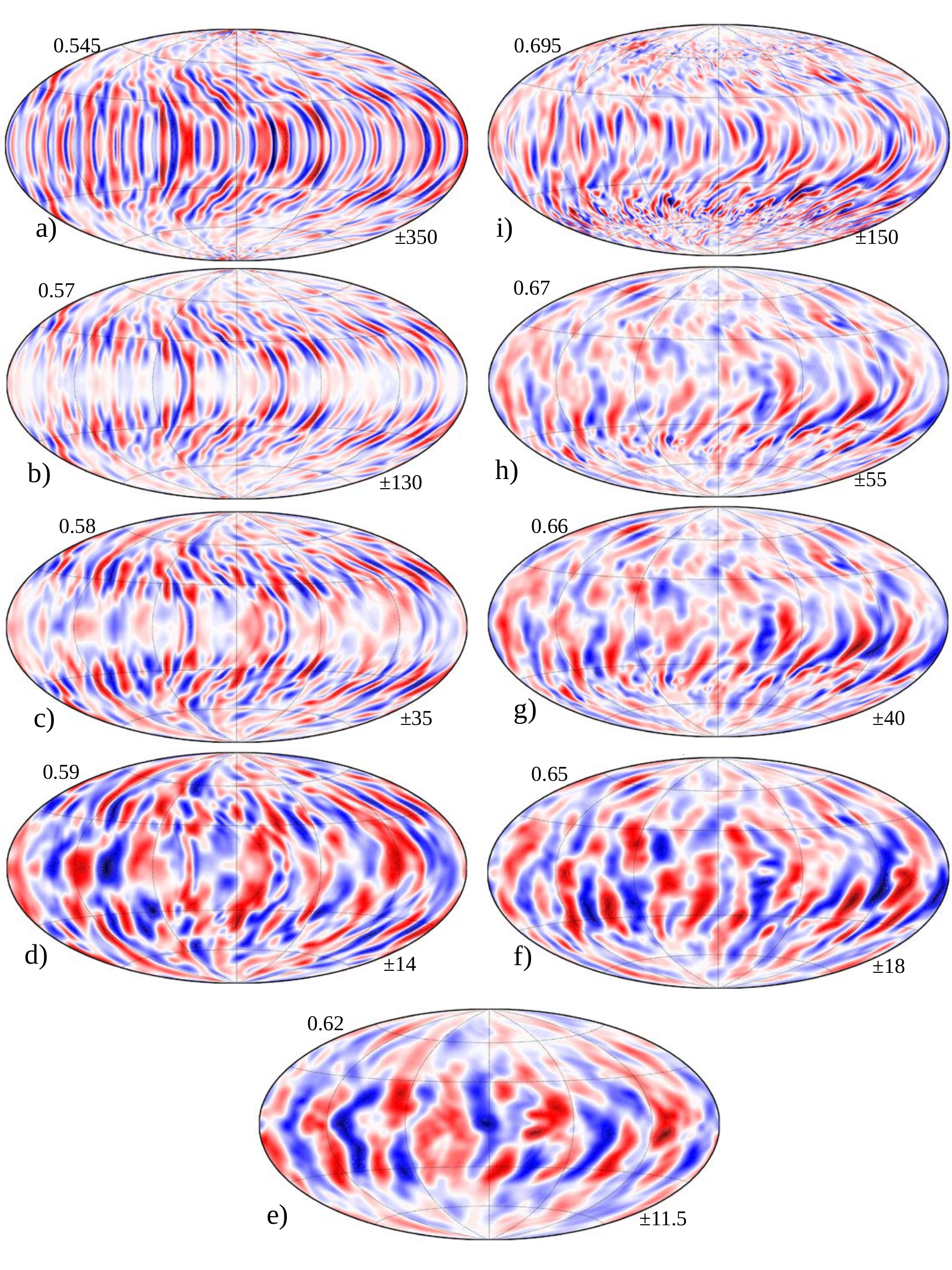}
\caption{Instantaneous radial flow on spherical surfaces taken at various radii indicated at the top left of each panel, the min/max velocity contour \revb{is indicated} at the lower right. The velocity scale is  the Reynolds-number. Parameters: model 4.10 from tab.~\ref{tabruns}, $I_s=0.746$, $Ro_c=0.245$.} 
\label{figconvssl}
\end{figure*}

The spherical surface projections (fig.~\ref{figconvssl}, a-i) reveal more details of the damping mechanisms. The nine maps represent several radial levels ranging from below the SSL (a), to the center $r/r_o=0.62$ (e), and on top of the SSL at $r/r_o=0.695$ (i). The typical columnar structures in the lower convective layer \revb{are apparent} outside TC (a, b). As those are more rotationally constrained by the Coriolis force they can penetrate deeper into the stable layer than the spiraling structures at higher latitudes and though cause the radial flow peaks around $\pm 60^\circ$ colatitude (b, c). Deeper inside the SSL, in the equatorial regions, large length scale flows appear more dominant (c), even exceeding the remaining columnar patterns found in higher latitudes (d, e). The rms radial flow amplitude \revb{decreases} drastically towards the center of the SSL, but at unequal rates for different colatitudes and different length scales. In the center of the SSL (fig.~\ref{figconvssl}, e) the flow is of weak amplitude and apparently dominated by the larger length scale. The amplitude has been decreased by a factor of 35, where the strongest remaining flows are concentrated in a broad belt around the equator. This is unlikely linked to rotational penetration as no columns can be vertically extended. \rev{For inertia penetration to be efficient, the flow must be energetic, what does not seem to be the case for those large scale and weak amplitude flows.} Towards the outer edge of the SSL (f-g), the flow amplitudes keep increasing featuring columnar elongated structures outside and small-scale, spiraling convection patterns inside the effective tangent cylinder (h-i). The effective TC is now attached to the upper edge of the SSL at $r_{ub}$, hence the columnar flows are confined to a much smaller colatitude range. 

This indicates that when columns touch on the stratified layer, they penetrate deeper, likely due to the rotational penetration or the Taylor-Proudman theorem. However, in the inner equatorial region and inside the inner TC, this effect is secondary. Even more so for the penetration at the outer boundary, where columns may only directly touch the stratified layer at a low-latitude band. As a result, the inertia penetration dominates, but is somewhat modified by latitudinal effects.

\begin{figure*}
\centering
\includegraphics[width=\textwidth]{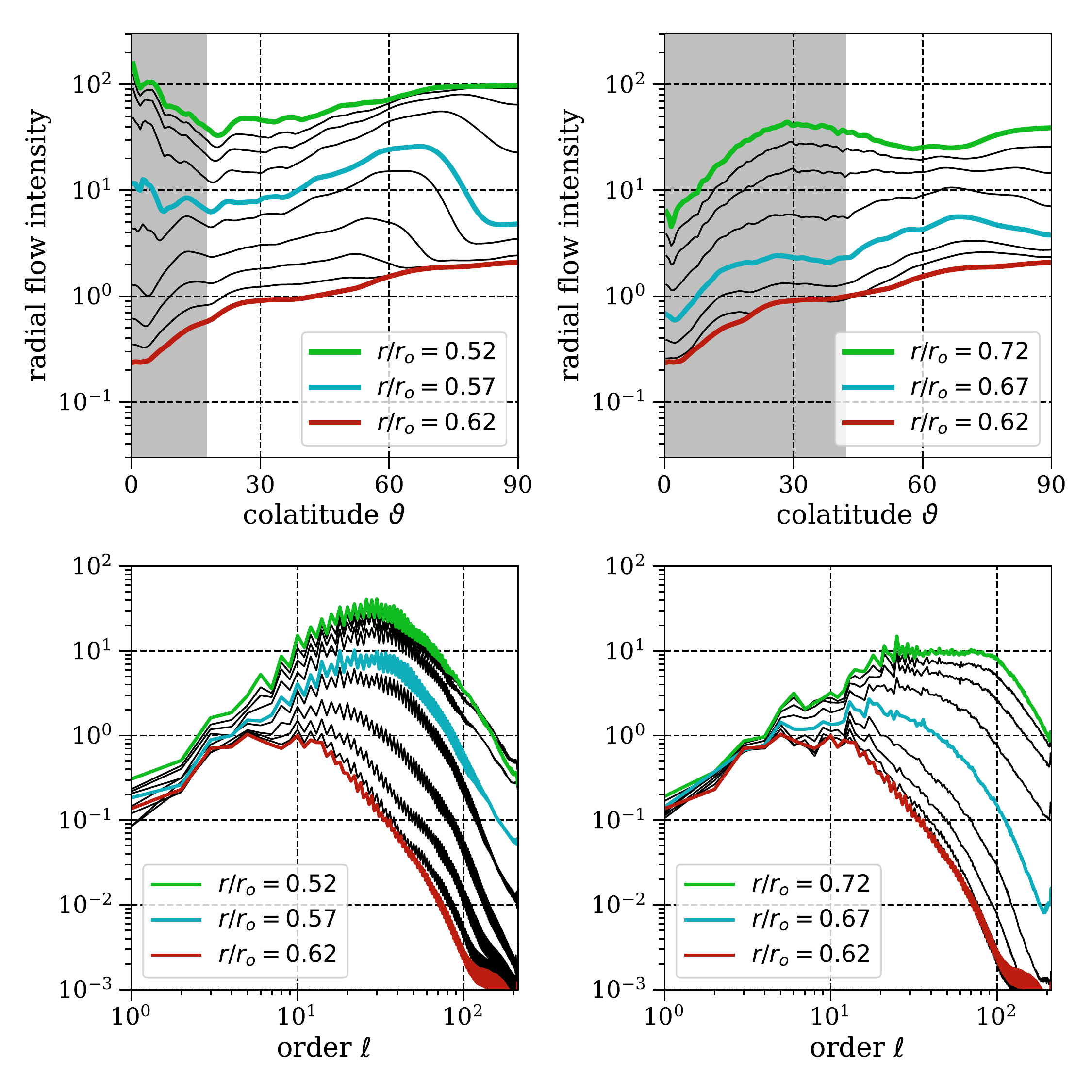}
\caption{Time and azimuthally averaged intensity of radial flows at various depth around the edge of the SSL $r_{lb}=0.57$ (left) and $r_{ub}=0.67$ (right). The upper plot show profiles along colatitude, the lower for spectral degree (length scale). \rev{The grey shaded areas mark colatitude bands inside the tangent cylinder set in the inner convective shell by $\sin \vartheta_{TC} = r_{i}/r_{lb}$ (top left) and in the outer convective shell by $\sin \vartheta_{TC} = r_{ub}/r_{o}$. } }
\label{figpradialprof}
\end{figure*}

\rev{For a more detailed analysis of the apparent colatitudinal variation and the length scales dependencies, a FFT transforms from $v_r(r,\vartheta)$  to $v_r^\star(r,\ell) $, where $\ell$ is the spherical harmonic degree}. For a few radii\revb{,} this is shown as a function of either colatitude or spectral degree in fig.~\ref{figpradialprof}. The radial flow intensity drops slightly before reaching the lower SSL edge at $r_{lb}/r_o=0.57$ (top left), though with weak modulation along colatitude $\vartheta$. \revb{At} the edge (blue line), the damping is strongest close to the rotation axis and around the equatorial region. In between rotational penetration somewhat reduces the damping of radial flows (profiles are closer to each other). This is the expected behavior, as the Coriolis forces act to extend convective columns into the stratified region predominantly at mid-latitudes outside TC. For the upper edge (top left), the region inside TC is larger and spans roughly \revb{the upper} half of \revb{the} colatitude \revb{domain}. There the radial flows decrease strongly with colatitude towards the poles, but are rather constant outside TC. In general, the radial flows close the equator remain strongest in the center of the SSL.
 
In the bottom plot of fig.~\ref{figpradialprof}, the radial flow per spectral degree $\ell$ is shown at the same depth levels used before. The broad peaks around $\ell=10-50$, indicate the mean convective horizontal length scales outside the SSL (green profiles). In this representation, it is obvious that small length scale flows are stronger suppressed by the inverse buoyancy gradients. Whereas at the short length scale end of the spectrum at $\ell \approx 100$ the total drop exceeds two orders of magnitude, the large length scales are damped less than one order of magnitude. Especially for the intermediate length scales around $\ell \approx 5-50$ the suppression seems most efficient. Comparing the spectra outside the SSL (green profile, fig.~\ref{figpradialprof}) with those at the bottom of SSL (red) clarifies why the flows seen before in fig.~\ref{figconvssl}, panel e) are so large scale. This behavior appears rather similar for both edges of the SSL (\revb{bottom} left). The effects of viscosity are only visible at the largest length scales, where the spectra follow different slopes. \rev{At the largest length scales, the flow amplitudes are not damped at all, but \revb{rather} maintained across the stable layer. This points towards an additional \revb{source} of radial flows powered by the convergence and dissipation of horizontal flows, which \revb{manifest} as  gravito-\revb{inertial} waves. }

\subsection{penetration depth}
\label{secpendepth}
\rev{For a qualitative assessment we assume that the penetration depth of radial flows is a linear function of the radial flow intensity itself \citep{Takehiro2001} and \revb{potentially} varying along colatitude. This implies, that  }
\begin{equation}
v_r(r,\vartheta) \propto \exp^{-r/\delta}\ ,
\end{equation}
where the penetration depth $\delta$  is defined by the e-fold decay scale height
\begin{equation}
\delta(r,\vartheta) = -v_r(r,\vartheta) \left[ \frac{\partial v_r(r,\vartheta)}{\partial r}\right]^{-1} \rev{= - \left[  \partial_r \log{v_r(r,\vartheta)}\right]^{-1}} \ .
\label{eqpendepth}
\end{equation} 
\rev{Fig.~\ref{figprofiles}, panel b) show that the radial profiles $v_r$ drops exponentially in the vicinity of the neutral buoyancy radii located at $r_{lb}/r_o=0.57$ and $r_{ub}/r_o=0.67$, respectively}. The penetration depth is calculated \rev{as a function of radius using eq.~\ref{eqpendepth}  while averaging over the colatitude $\vartheta$
\begin{equation}
\delta_r(r) = - \left[  \partial_r \log{\tilde{v}_r(r)}\right]^{-1} \qquad \text{with} \qquad \tilde{v}_r = \frac{1}{2} \int_0^\pi v_r(r,\vartheta) \sin \theta d\theta \ 
\end{equation}
or as function of colatitude taken at the lower edge of the stratified region 
\begin{equation}
\delta_\vartheta (\vartheta) = - \left[  \partial_r \log{v_r(r_{lb},\vartheta)}\right]^{-1} \ .
\end{equation}
Finally\revb{,} we also investigate the length scale dependence in terms of the spherical harmonic degree $\ell$ as used in the linear theory study by \citet{Takehiro2001}:
\begin{equation}
\delta_\ell (\ell) = - \left[  \partial_r \log{v^\star_r(r_{lb},\ell)}\right]^{-1} \ ,
\end{equation}
where $v^\star_r(r,\ell) = \operatorname{FFT}\left\{v_r(r,\vartheta) \right\}$ is the Fourier-transformed.}
\begin{figure*}
\centering
\includegraphics[width=0.9\textwidth]{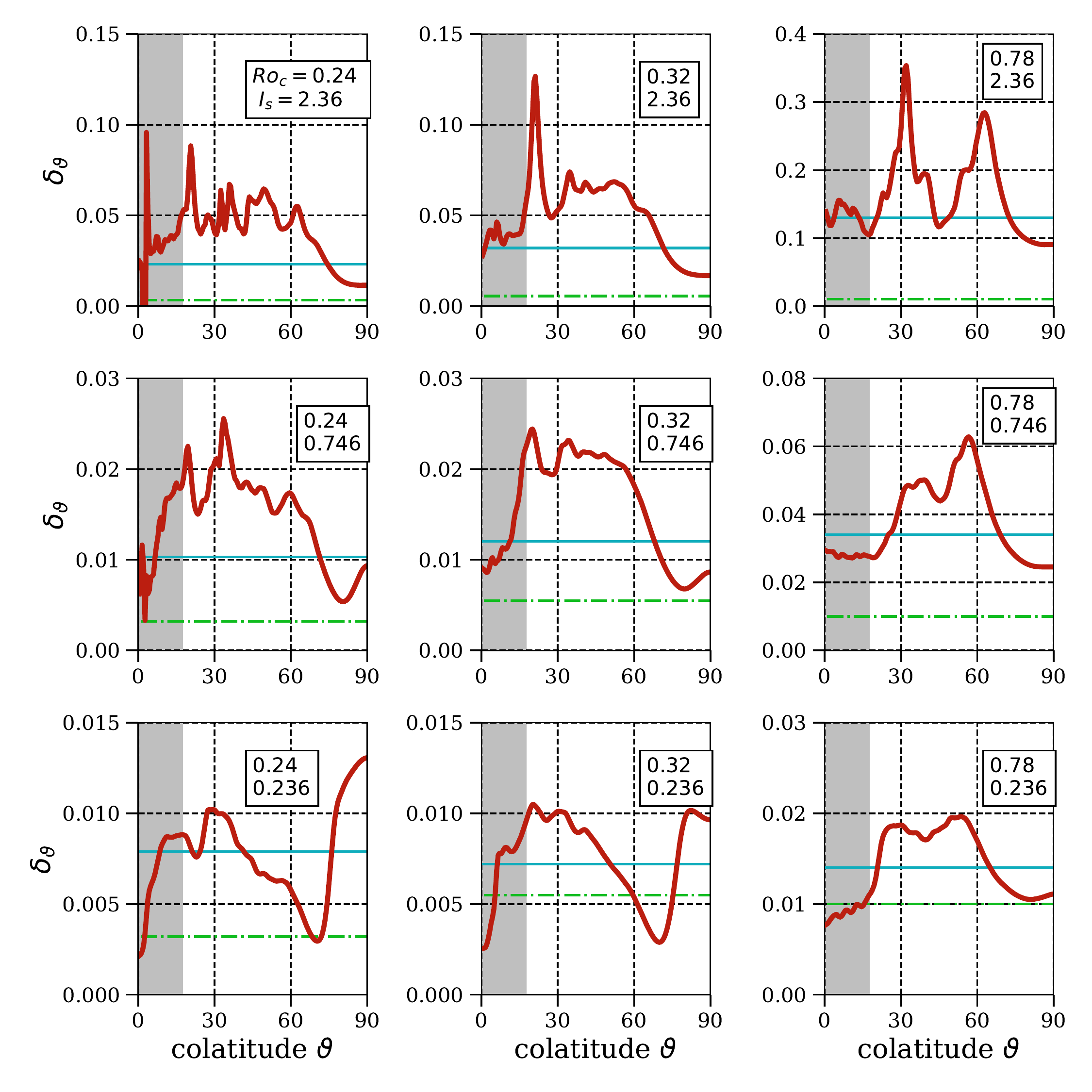}
\caption{Penetration depth of radial flows as a function of colatitude for nine cases from tab.~\ref{tabruns}, group 4 taken at the lower edge of the SSL $r=r_{lb}$. The green  line shows the thickness of an Ekman layer $\delta_\nu=\sqrt{E}$. \rev{The grey shaded areas mark colatitude bands inside the tangent cylinder set by $\sin \vartheta_{TC} = r_{i}/r_{lb}$.} }
\label{figpendepth_colat}
\end{figure*}

\revb{Fig.~\ref{figpendepth_colat} shows the} colatitude dependence of $\delta_\vartheta$ for \revb{the} nine cases (group 4 in tab.\ref{tabruns}). The reference model \revb{(no. 4.10)} previously being subject of a detailed investigation, is placed left \revb{in the} middle row. The other cases are set up such that $Ro_c$ is \revb{constant} for each column and $I_s$ for each row (both indicated at the top right). Inertia increases from left to right, the stratification strength relative to the rotation rate from top to bottom. 

The overall penetration depth decreases with decreasing inertia ($Ro_c$) and increasing stratification ($I_s$). For the top right model, the inertia is strong and the stratification weak ($Ro_c =0.78$, $I_s=2.36$), such that the penetration depth exceeds significantly the thickness of the SSL ($0.1\,r_o$). If $I_s$ is ten-fold decreased (bottom right) the penetration depth is reduced to a fraction of the SSL width \revb{and} the radial flows are \revb{hence} tiny deep inside the SSL. For the other columns, smaller Ekman numbers provide smaller $Ro_c$ and consequently smaller penetration depth as the efficiency of overshooting convection is reduced. All models seem to share significant higher penetration depths at mid-latitudes. Inside the TC and in equatorial regions the penetration depth is typically smaller. However, the penetration depth is bounded by the viscosity (horizontal green lines, $\delta_\nu=\sqrt{E}$) and apparently \revb{cannot} be reduced far below this value. We conclude that the penetration depth at intermediate colatitudes \revb{of} $30^\circ - 60^\circ$ can easily reach twice its mean value, indicated by the blue lines, what is most clearly seen in the bottom right panel. It was shown before that large scale flows remain \revb{most dominant} in the equatorial regions. Those are obviously only weakly damped (fig.~\ref{figpradialprof}, bottom\revb{,} yield larger penetration depth in the equatorial regions (fig.~\ref{figpendepth_colat}, left bottom) \revb{and} might originate from collision of horizontal flows. 

\begin{figure*}
\centering
\includegraphics[width=0.9\textwidth]{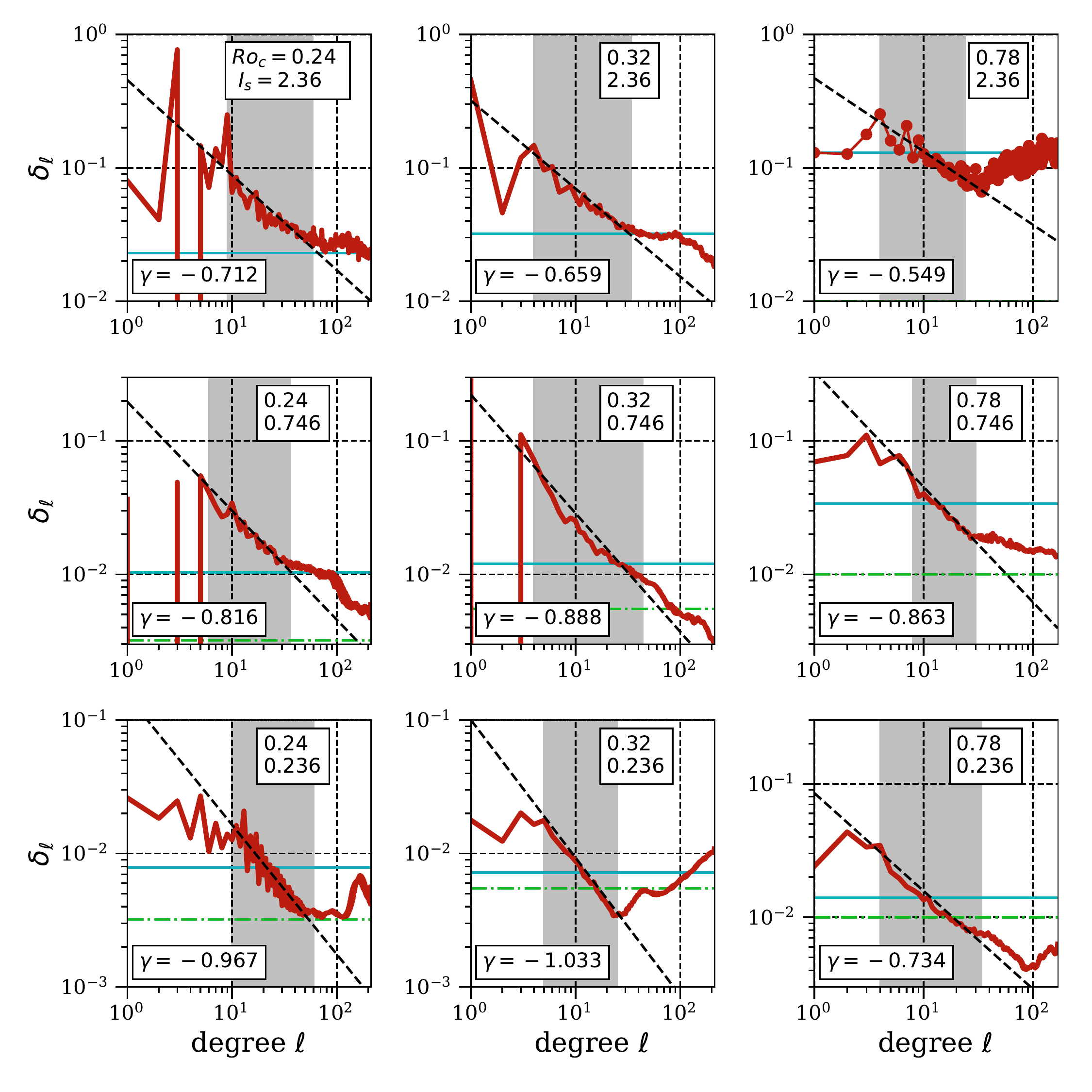}
\caption{Penetration depth as a function of spectral degree $\ell$ for the nine case of group 4 in tab.~\ref{tabruns}. The horizontal blue lines indicate the mean values. The green line shows the thickness of an Ekman layer $\delta_\nu=\sqrt{E}$. The black lines show power laws fitted over the grey-shaded length scale ranges resulting into a decay exponent $\gamma$ added at the bottom of each figure.}
\label{figpendepth_spec}
\end{figure*}

\citet{Takehiro2001} concluded that the stable layer acted as a spectral filter with \revb{low pass} nature, more precisely they predict that $\delta_\ell$ should decay like $1/\vert \ell \vert $ . Fig.~\ref{figpendepth_spec} shows $\delta_\ell$ for the nine cases from group 4 (see tab.~\ref{tabruns} and fig.~\ref{figpendepth_colat}). The blue horizontal lines denote the mean value listed in tab.~\ref{tabruns}, the green one marks the thickness of the Ekman layer. 

It can be seen, that penetration depth clearly decays for smaller length scales even for cases with rather large inertia and weak stratification (top right). As shown for the case 4.10 in fig.~\ref{figpradialprof}, the largest (small $\ell$) and smallest length scales (large $\ell$) contain little kinetic energy and are excluded from the analysis. Therefore\revb{,} power laws of the form
\begin{equation}
\delta_\ell \propto \ell^\gamma
\end{equation}
are fitted over an intermediate $\ell$-range \revb{that is} highlighted in grey. The resulting spectral decay exponent is added on the bottom of each panel. The slopes are all negative and spread amongst $\gamma = [-0.55,-1.03]$, where models with weak inertia and strong stratification approach the $(\gamma=-1$)-scaling predicted by the linear theory of \citet{Takehiro2001}. Those models are most effectively controlled by rotational penetration and not overly biased by inertia. This indicates that the results of \citet{Takehiro2001} can only be applied to an intermediate length scale range, for which the agreement is striking. The large $\ell$ part, which has been excluded \revb{from the fit}, might be controlled by the viscosity and hence the penetration depth is not decreasing below the \revb{thickness of the} Ekman layer. For the small $\ell$ part of the spectrum\revb{,} the associated flows are controlled by other means, where the large scale patterns found in the equatorial regions (compare fig.~\ref{figconvssl}, panel e) or meridional circulation induced by the latitude-dependent penetration might play a role. Also other \revb{typical flows} found in stratified regions, such as \revb{gravito-inertial} waves could be taken into account. Those \revb{ might then be} the same features seen before in the equatorial regions (fig.~\ref{figconvssl},e ) deep inside the SSL around the equator. \revb{If applicable, the radial waves are then excited by dissipating or diverging the horizontal counterparts preferentially in the equatorial belt.}

\begin{figure}
\centering
\includegraphics[width=0.6\columnwidth]{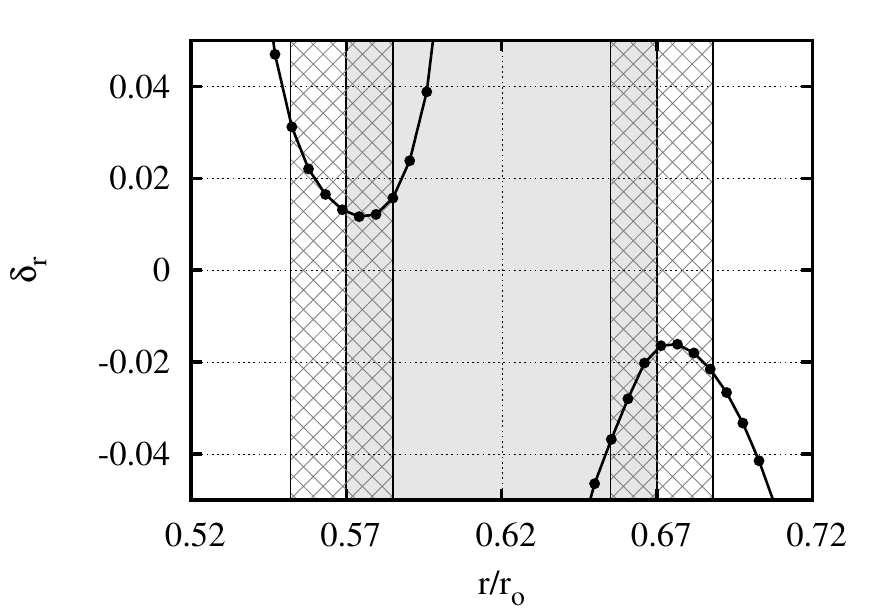}
\caption{Mean penetration depth along radius for model 4.10. The grey shaded area marks the SSL\revb{;} the hatched areas indicate the radius range in which the extrema of $\delta_r$ are taken.}
\label{figtest_pen}
\end{figure}

\begin{figure*}
\centering
\includegraphics[width=1.0\textwidth]{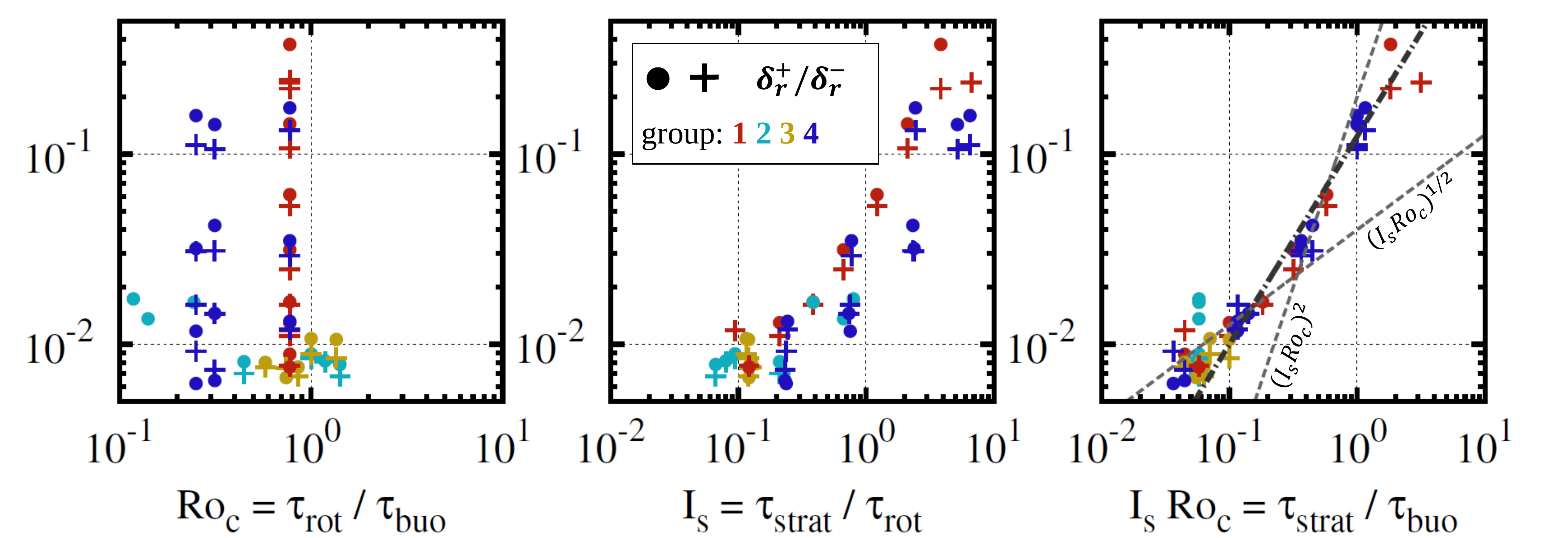}
\caption{Mean penetration depth at the lower edge ($\delta_r^+$, filled circles) and upper edge ($\delta_r^-$, plus markers) as a function of time scale ratios. Plotted are models from tab.~\ref{tabruns}, colors indicate various groups. For the right hand plot a power law is fitted to the data for $I_s\, Ro_c > 0.1$ \rev{(thick black, dot-dashed line)\revb{,}} showing the \revb{almost} linear dependence of $\delta_r \propto I_s\,Ro_c$. \rev{The thin grey \revb{power laws} indicate previously published scaling suggestions. For details\revb{,} see text.} }
\label{figpendepth_mean}
\end{figure*}

Finally\revb{,} the dependence of the mean penetration depth on the model parameters ($Ra$, $E$, $Pr$ and $A_{SSL}$) is studied. The penetration depth at the lower and upper boundary of the SSL are indicated by $\delta_r^+$ and $\delta_r^-$. Those are derived by taking the minimal value of $\delta_r$ in the vicinity ($\pm 0.01r_o$) of $r_{lb}$ and the maximal value of $-\delta_r$ \revb{near} $r_{ub}$ (see also fig.~\ref{figtest_pen}). 

Fig.~\ref{figpendepth_mean} shows $\delta_r^+$ and $\delta_r^-$ for all models from tab.~\ref{tabruns}. It can be clearly seen, that $\delta_r$ decreases with increasing stratification and settles somewhat below $\delta_r=0.01$. This lower boundary is due to viscosity and hence depending on the Ekman number. For all studied Ekman numbers $\delta_r$ can be decreased to values \revb{approximatively the Ekman layer thickness }  \revb($\delta_E = E^{1/2}$) suggesting that our numerical setup is only limited by the viscosity. \rev{Models with $\delta_r$ larger than the width of SSL ($0.1\,r_o$) are robustly constrained, as the mean radial flow intensity drops noticeable across the SSL, yet less than e-fold. As shown by the dark-blue profile in the left panel of fig.~\ref{figprofiles}, even models with neutral buoyancy gradient ($A_{SSL}=0$) in the SSL create a strong dimple in the poloidal energy.} Apparently $\delta_r^+$ and $\delta_r^-$ are nearly identical, suggesting that there is no major dynamic difference between a scenario where the stable region is on top of the convective one and the opposite situation. As the \revb{Coriolis force clearly dominates the} flows in convective regions ($Ro_c < 1$), it appears reasonable to expect that $I_s = \tau_{strat} / \tau_{rot}$ is the appropriate force ratio to estimate the penetration depth (fig.~\ref{figpendepth_mean}, left). However, this does not seem to hold for all Ekman numbers. The right hand plot of fig.~\ref{figpendepth_mean} shows an alternative scaling attempt relative to $I_s \, Ro_c = \tau_{strat} / \tau_{buo}$, i.e.~ the rotational time scale is replaced with the inertia time scale. This indicates that the penetration depth $\delta_r$ is independent of rotation and depends, as in a non-rotating system, only on the ratio of inertia and stratification. Further\revb{,} a power law fit of the form  
\begin{equation}
\delta_r = a \cdot  (I_s\,Ro_c )^b 
\end{equation}
\revb{results} in $a = 0.121 \pm 0.008$ and $b=0.989 \pm 0.1162$ and hence suggesting a linear dependence. For the \revb{least-squares} fitting \revb{procedure} only data points were considered for which $I_s\, Ro_c > 0.1$, since they are not biased by viscosity \revb{(thick black, dot-dashed line in fig.~\ref{figpendepth_mean}, right panel)}.

\rev{Previous studies targeted to the radiative-convective boundary in the Sun, such as \citet{Hurlburt1994,Brummell2002,Rogers2006} controlled the degree of stratification by reducing or increasing the local polytropic index $n_i$ below or above the adiabatic one $n_a$. Using the polytropic relations \revb{(e.g. $\rho = \rho_o \xi^n$, where $\xi$ is a function of $r$)}, the deviation from the adiabat can be expressed by logarithmic density gradients and polytropic indices
\begin{equation}\frac{1}{\rho} \frac{d\rho}{dr} - \frac{1}{\rho} \frac{d\rho}{dr} \Big\vert_a \propto \left( n_i -n_a\right) \frac{\xi^\prime}{\xi }\propto \frac{1}{c_p} \frac{ds}{dr} \ ,
\end{equation}
where index $a$ refers to the adiabat. \revb{Consequently}, the stratification measure $S$ defines the ratio of subadiabatic (positive) to superadiabatic (negative) entropy gradients and is compatible with our stratification amplitude $S \approx A_{SSL}$. Whereas the numerical models in the previous study cover stratifications between $S=1$ and $S=30$, our parameter space with respect to $A_{SSL}$ spans from 0 to 750 (see tab.~\ref{tabruns}).  Further the scaling above suggested that 
\begin{equation}
\delta_r \propto I_s Ro_c = 2 \left(  g(r) \frac{d\bs{s}}{dr} \right)^{-1/2} \approx A_{SSL}^{-1/2} \ .
\end{equation}
Hence\revb{,} our results show that the penetration depth increases with the square root of the ratio of positive and negative entropy gradients. Firstly reported by \citep{Hurlburt1994}, comparable models found different scaling exponents, which changed as a function of $S$.  Typically for small $S$ the penetration depth scales with $\delta_r \propto S^{-1}$, but for larger S with a significantly shallower $S^{-1/4}$ \citep{Hurlburt1994}. Other studies only found the weak scaling throughout the studied $S$-range \citep{Brummell2002, Rogers2006}. \revb{For guidance, both} scaling \revb{laws} are \revb{drawn} into fig.~\ref{figpendepth_mean}, right panel, \revb{thin grey dashed lines}. Indeed those \revb{predictions} seem to approximate the data in local sub-regions \revb{of the parameter space}. However, the general trend appears much  better reproduced by a single scaling law with an exponent of roughly $S^{-1/2}$. }

\section{summary and discussion}
We have performed an extensive numerical modeling campaign of rapidly rotating convection in a spherical shell \revb{that exhibits a sandwiched stable stratified layer between two convective zones}. Such a system is most suitable for Saturn, where the H/He demixing generates a compositional gradient around mid-depth and though \revb{suppresses} thermal convection locally. Our numerical models cover a large fraction of the appropriate parameter space, e.g. in the terms of the stratification relative to rotational forces $2\Omega/N$, where $N$ is the Brunt-V\"ais\"al\"a frequency or relative to the inertia characterized by $Ro_c$ (see also tab.~\ref{tabruns}).

Stratification acts directly to suppress radial flows, but radial flows can pierce across the interface by different mechanisms. The classic overshooting penetration depends on \revb{the} inertia a fluid parcel has gained from buoyancy instabilities. \revb{Furthermore}, rapid rotation and the spherical geometry allows convective columns to be vertically extended into the stratified regions \revb{in accordance with the Taylor-Proudman theorem}. We investigate and distinguish the potential effects of both penetration mechanisms by altering the leading order force balances. \revb{Our results indicate} that the mean penetration depth depends linearly on the square root of the ratio of \revb{destabilizing} and \revb{stabilizing} entropy gradients ($\delta_r \approx A_{SSL}^{-1/2}$). \rev{This builds on the rather diverse results derived in the solar context \citep{Zahn1991,Hurlburt1994, Brummell2002,Rogers2006}, where different scaling exponents with stronger scaling for weaker stratification were reported. The \revb{exponents} in those studies range from $-1/4$ to $-1$, but the robust $-1/2$-scaling suggested in the present study reproduces the results convincingly. \revb{In addition,} and in contrast to the expectations from \citet{Takehiro2001}, the \revb{magnitude of the} penetration depth appears largely unaffected by the rotation.} 
 
The Coriolis force, however, adds a characteristic latitudinal modulation of the penetration depth amplifying to $\pm (40-50)\%$ of the mean value, where the strongest penetration resides at mid-latitudes (outside TC). This clearly reflects the action of columnar structures vertically extended by the Coriolis force and hence \revb{represents} rotational penetration. \revb{Therefore}, latitudinal density gradients are expected to \revb{build} up at the edges of stratified regions. Such gradients are known to drive baroclinic instabilities, e.g. thermal winds. 

Alongside original studies using linear waves, such as \citet{Takehiro2001}, a spectral \revb{(length scale)} dependence of the penetration depth is \revb{expected} and this \revb{is indeed} found in our strongly non-linear simulations. The decay exponent $\gamma$ approaches unity, what confirms the expected value from linear theory \citep{Takehiro2001}. The SSL though clearly shows the quality of a \revb{low pass} filter. Consequently, radial flow in the SSL\revb{,} even though weak in amplitude\revb{,} must be expected to be of larger length scale than the convective flows pummeling into the SSL. However, the striking agreement with the theoretical predictions is limited \revb{to} an intermediate length scale range. 

As a peculiar result, we have identified \revb{wave-like radial flows that seem to somewhat resist the inverse buoyancy forces in the stratified layer. Those are non-geostrophic, of large \revb{length scale}, weak \revb{in} amplitude and \revb{predominantly} driven in the equatorial regions of the stratified layer}. \rev{We showed that the required additional source of radial flow is likely not due to \revb{the} length scale \revb{filtering effect of the rotational penetration} or due to inertia penetration, but \revb{is potentially} powered from the convergence of horizontal flows \revb{in the shape of} gravito-\revb{inertial} waves.}

\revb{Finally,} we have shown that viscous effects only play a role for very shallow penetration depth and at the smallest length scales. The overall results should not depend on the fact that we have only explored an Ekman number rage limited by the numerical constraints. More important is the spherical geometry. We speculate that a thin stable stratified layer will be significantly \revb{penetrated by Coriolis force induced columnar flows vertically extending into the SSL} and hence more closely obey the linear theory by \citet{Takehiro2001}. 

As an outlook for future work, a thorough investigation of the gravito-inertial waves generated in the SSL, \rev{the emergence of thermal winds due to latitude-dependent penetrative convection} and consequently how differential rotation is paused through stratified zones. \revb{In addition,} the effect on a dynamo process originating from underneath will help to understand the dynamics and \revb{magnetic fields} of partly stratified rapidly rotating convection, which is most suitable for the planetary atmospheres and interiors.

\section*{Acknowledgments}
\rev{The authors would like to thank the anonymous reviewers for their valuable comments and suggestions to improve the manuscript.} This work was supported by the Deutsche Forschungsgemeinschaft (DFG) in the framework of the priority programs SPP 1488 'Planetary Magnetism' and SPP 1992 'Diversity of Exoplanets'. MagIC is available at an online repository (https://github.com/magic-sph/magic).

\section*{Conflict of Interest Statement}
The authors declare that the research was conducted in the absence of any commercial or financial relationships that could be construed as a potential conflict of interest.

\section*{Author Contributions}
WD designed and carried out the numerical experiments, analyzed the data and wrote the paper manuscript. JW provided physical insights, fluid dynamics expertise and significantly contributed to the manuscript.

\bibliographystyle{frontiersinSCNS_ENG_HUMS} 
\bibliography{pendepth}

\end{document}